\documentclass[11pt]{article}

\usepackage{graphicx}
\usepackage{amsmath}
\usepackage{amssymb}
\usepackage[authoryear]{natbib}
\usepackage{xcolor}
\usepackage{soul}
\usepackage{longtable}
\usepackage{booktabs}
\usepackage{array}
\usepackage[margin=1in]{geometry}
\usepackage{placeins}
\usepackage[
  colorlinks=true,
  linkcolor=blue,
  citecolor=blue,
  urlcolor=blue
]{hyperref}
\usepackage{doi}
\usepackage{cleveref}
\usepackage{makecell}
\usepackage{rotating}

\title{Projections of Earth's Technosphere: \\ Civilization Collapse--Recovery Dynamics and Detectability}
\author{\begin{tabular}[t]{c}Celia Blanco$^{1,2}$, Jacob Haqq-Misra$^{2}$, George Profitiliotis$^{2}$\\[0.5em]
\small{1. Centro de Astrobiolog\'{i}a (CSIC-INTA),}\\
\small{Ctra de Torrej\'{o}n a Ajalvir, km 4, 28850 Torrej\'{o}n de Ardoz, Madrid, Spain}\\[0.3em]
\small{2. Blue Marble Space Institute of Science, Seattle, WA 98104 USA}\\[0.5em]
\small{Corresponding author: \href{mailto:celia.blanco@cab.inta-csic.es}{celia.blanco@cab.inta-csic.es}}\end{tabular}}
\date{}

\begin{document}

\maketitle

\section{Abstract}\label{abstract}

How long a technological civilization remains active, and what determines whether it collapses or persists, is a central question for both projecting humanity's future and assessing the prevalence of detectable intelligence in the galaxy. We model collapse-recovery dynamics across ten plausible futures for Earth-originating civilization using a hybrid deterministic-stochastic simulation over a 1000-year window. The duty cycle, defined as the fraction of its total lifespan that a civilization is technologically active, ranges from $\sim$0.38 to 1.00, with trajectory outcomes shaped by the interplay of governance structure, resource pressure, and hazard exposure. Several model parameters map onto actionable resilience levers, and modest improvements can qualitatively alter long-term trajectories. Sensitivity analysis reveals that the resource depletion rate and the post-collapse recovery fraction are consistently the most impactful levers across scenarios, suggesting that reducing resource consumption may be at least as important as mitigating existential hazards for avoiding civilizational collapse. We discuss implications for Earth's civilizational resilience and for the search for extraterrestrial technosignatures. We also derive an effective detectability duration that accounts for intermittent civilizational activity, and show that the apparent absence of extraterrestrial signals may reflect the prevalence of low-duty-cycle civilizations rather than the rarity of intelligent life.

\vspace{1em}
\noindent\textbf{Keywords:} civilizational collapse; technological resilience; duty cycle; technosignatures; existential risk

\section{Introduction}\label{introduction}

One of the most puzzling questions in astrobiology is whether intelligent civilizations exist elsewhere in the galaxy, and if so, why we have not yet detected them. This apparent contradiction, often referred to as the Fermi Paradox or the Great Silence, raises the question of why, in a galaxy billions of years old, we observe no clear signs of advanced extraterrestrial life. Many solutions have been proposed to explain this Great Silence \citep{Webb2015-vi}. One class of explanations suggests that intelligent life is intrinsically rare, perhaps requiring an improbable convergence of planetary, chemical, or biological conditions \citep{Wesson1990-xi}. Another contends that while intelligence may emerge, technological civilizations are typically short-lived, succumbing to self-destruction through ecological collapse, conflict, or other systemic instabilities \citep{Vinn2024-pi}.

The second class of explanation raises a question that is critical regardless of how common the emergence of intelligence may be: how long do technological civilizations last once they arise? Even if intelligent life is widespread, the number of civilizations detectable at any given time depends decisively on their longevity \citep{Balbi2021-bc}. This question is also central to projecting the long-term trajectory of our own civilization. Most treatments of longevity, however, assume that once a society becomes technologically active, it remains so continuously until it either collapses or deliberately goes silent. But this assumption is difficult to justify. Historical studies of Earth's civilizations reveal cycles of rise and fall with typical durations on the order of centuries \citep{Shermer2002-oi}, and even within a single civilization's lifespan, the temporal distribution of active periods may matter as much as their cumulative duration \citep{Balbi2018-fp}. Therefore, understanding civilizational longevity requires moving beyond static estimates and examining the temporal dynamics of technological activity itself.

There are good reasons to expect that civilizational collapse-recovery dynamics are complex. It has been argued that planetary civilizations face a fundamental bifurcation: either "asymptotic burnout" in which superlinear growth leads to singularity-driven collapse, or "homeostatic awakening", in which a civilization deliberately curbs expansion to achieve long-term equilibrium \citep{Wong2022-hv}. Sustainability constraints make exponential expansion highly unlikely, so advanced societies may stabilize or collapse before spreading across the Galaxy \citep{Haqq-Misra2009-jb,mullan2019population}. If collapse is common and recovery uncertain, then the technosphere (the totality of a civilization's technological infrastructure and output) may best be understood not as a permanent state but as an intermittent phenomenon, alternating between periods of activity and dormancy in a pattern shaped by resource availability, governance structure, and exposure to existential risk.

Technological intermittency can be quantified through a civilization's \emph{duty cycle}, defined as the fraction of its total lifespan during which it is technologically active. The duty cycle captures a reality that static longevity estimates miss---that a civilization persisting for a thousand years may spend substantial portions of that time in dormancy, recovery, or regression. In the context of the search for extraterrestrial intelligence, the relevance of intermittent activity is already recognized. Interstellar signals may be intermittent due to power constraints, planetary rotation, or targeted sequential transmissions \citep{Gray2020-cn}, and recent work applying Lindy's law to technosignature longevities suggests that technologically active periods are more likely to be short-lived than long-lived \citep{Balbi2024-bg}. But the implications of intermittency extend well beyond detectability. For our own civilization, understanding the dynamics of collapse and recovery, and how governance, resources, and resilience infrastructure shape the duty cycle, is directly relevant to long-term planning and existential risk mitigation.

In this paper, we develop a framework for modeling these dynamics. Building on a recent scenario framework for Earth-originating civilization \citep{Haqq-Misra2025-rt}, we simulate collapse-recovery cycles across ten diverse sociotechnical trajectories spanning 1000 years and compute duty cycles for each. We then derive an effective detectability duration, $L_{\mathrm{eff}}$, and explore the implications for Earth's own civilizational resilience, for the Drake Equation's longevity term \emph{L}, and for the broader search for technosignatures. 

\section{Model Description}\label{model-description}

A recently proposed framework outlines ten plausible futures for Earth-originating civilization 1000 years from now, structured around variations in governance type, resource regime, and systemic resilience \citep{Haqq-Misra2025-rt}. These scenarios are grounded in Earth's present-day sociotechnical conditions and extrapolated using techniques from futures studies, including PEST analysis (Political, Economic, Social, Technological), cross-consistency assessment, and structured scenario worldbuilding. Building on this qualitative scenario set, we implement a hybrid deterministic-stochastic simulation to represent the temporal dynamics of growth, collapse, and recovery.

\subsection{Simulation framework}\label{simulation-framework}

The simulation spans a fixed period of 1000 years, discretized into yearly time steps. At each step, a civilization is characterized by its technological capacity, \(T(t)\), and its available resource stock, \(R(t)\). Technological capacity increases linearly each year according to a fixed growth rate \(r\), which we take directly from the values reported in \citet{Haqq-Misra2025-rt}, Table 9. Simultaneously, the available resource stock is depleted at a constant rate \(\delta\). If the resource level falls to zero (i.e., \(R(t) < 0\)), or if an existential risk event occurs (represented as a probabilistic hazard drawn at each time step) the civilization undergoes a collapse. Collapse represents a systemic breakdown in infrastructure, coordination, and technological continuity. It leads to a sharp reduction in technological capacity, determined by a collapse factor \(c_{f}\), which defines the proportion of technology that survives the event. This loss can be interpreted as the result of disruptions to power grids, communication networks, knowledge retention, or the abandonment of complex systems.

Following collapse, the system enters a dormant recovery period of duration \(r_{d}\). During this phase, the society is unable to grow or consume resources, reflecting a temporary halt in technological and economic activity. This delay allows time for institutional reorganization, population stabilization, or environmental recovery. Once the recovery period concludes, the resource stock is partially replenished to a fraction \(r_{f}\) of its original value \(R_{0}\). This recovery fraction represents retained infrastructure, stored reserves, or ecological regeneration. A full recovery (\(r_{f} \approx 1\)) implies robust contingency planning and strong institutional memory, while lower values indicate more severe losses and diminished rebuilding capacity. This cycle of growth, collapse, dormancy, and recovery can repeat multiple times during the simulation window, depending on the fragility of the system, the frequency of stochastic hazards, and the depletion rate of resources.

Formally, let the initial conditions be \(T_{0} = 1\), and \(R_{0} \in {\mathbb{R}}_{+}\ \), and define the indicator function as \(\chi(t) \in \{ 0,1\}\), with \(\chi(t) = 1\) if the system is active (not collapsed or recovering) at time \(t\), and \(\chi(t) = 0\) otherwise. The update rules for each time step are given by
\begin{equation}
\begin{aligned}
T_{t+1} &= \chi_t \left( T_t + r \right) + \left(1 - \chi_t\right) T_t ,\\
R_{t+1} &= \chi_t \left( R_t - \delta \right) + \left(1 - \chi_t\right) R_t .
\end{aligned}
\label{eq:update}
\end{equation}

We include both a deterministic and a stochastic trigger for collapse to reflect two distinct failure modes: internal depletion and external shocks. Collapse is deterministically triggered when the resource stock is fully depleted (i.e.,\(\ R(t) \leq 0\)). It may also be triggered by a stochastic existential hazard, which we model as a Bernoulli process. At each time step \emph{t}, a uniform deviate \(U(t)\ \sim\ \ U(0,\ 1)\) is drawn. If \(U(t)\  < h\), where \(h \in \lbrack 0,1\rbrack\) is the hazard rate specific to the scenario, then a collapse is triggered independently of resource levels. 

Upon collapse, the system state is reset according to
\begin{equation}
\begin{aligned}
T_{t+1} &= c_f \, T_t ,\\
R_{t+1} &= 0 .
\end{aligned}
\label{eq:collapse}
\end{equation}

The system then remains inactive for $r_d$ time steps, after which resource availability is restored as
\begin{equation}
R_{t+r_d} = r_f \, R_0 .
\label{eq:recovery}
\end{equation}

Together, Equations~\eqref{eq:update}--\eqref{eq:recovery} define the full dynamical cycle of growth, collapse, and recovery in the model.

We define a civilization as being ON when it maintains a positive level of resources (i.e., \(\ R(t) > 0\)). This assumption reflects the idea that some minimal material throughput is necessary for societal function and detectability, including energy consumption, communication, and technosignature production. Conversely, when resources are fully depleted (\(R(t) = 0\)), the civilization is considered OFF, representing a dormant, collapsed, or otherwise undetectable state. This binary definition underlies the calculation of duty cycle and related metrics discussed in subsection ~\ref{metrics}.

\subsection{Scenario typologies and parameter mapping}\label{scenario-typologies-and-parameter-mapping}

We apply our model to the ten future scenarios to simulate the long-term behavior of civilizations under diverse sociotechnical conditions. Each scenario describes a distinct vision for humanity's trajectory, defined by combinations of resource abundance, governance structure, institutional resilience, and technological integration. Our goal is to translate these qualitative narratives into a coherent set of quantitative parameters that govern dynamics of growth, collapse, and recovery. Scenarios are characterized along three conceptual axes: resource regime (scarcity vs. non-scarcity), governance type (from autocratic to participatory), and structural design (hierarchical vs. distributed). Scenarios are also mapped onto six thematic clusters (Table~\ref{tab:scenarios}).

\begin{table*}[htbp]
\centering
\caption{Scenario descriptions and sociotechnical tags}
\label{tab:scenarios}
\setlength{\tabcolsep}{8pt}
\renewcommand{\arraystretch}{1.25}
{\raggedright
\begin{tabular}{l p{4.5cm} p{5.5cm}}
\toprule
Scenario & Tags & Description \\
\midrule
S1: Big Brother is Watching &
Scarcity + Rule by One + Hierarchical \newline Cluster 6 (Earth 2024) &
Centralized authoritarian system under resource stress \\

S2: Wild West &
Scarcity + Rule by Few + Hierarchical \newline Cluster 6 (Earth 2024) &
Deregulated, competitive, and fragmented society \\

S3: Golden Age &
Non-scarcity + Rule by All + Distributed \newline Cluster 6 (Earth 2024)$^{\dagger}$ &
Stable egalitarian society with abundant resources \\

S4: Living with the Land &
Scarcity + Rule by Few + Hierarchical \newline Cluster 5 (Simplicity) &
Low-impact, cyclical, ecologically constrained society \\

S5: Transhumanism &
Non-scarcity + Rule by All + Distributed \newline Cluster 4 (Engineering) &
Technologically integrated, post-biological civilization \\

S6: Sword of Damocles &
Scarcity + Rule by Few + Hierarchical \newline Cluster 4 (Engineering) &
High-tech and high-risk system with cascading failure potential \\

S7: Restoration &
Non-scarcity$^{\ddagger}$ + Rule by All + Distributed \newline Cluster 3 (Sustainability) &
Recovered post-collapse world with resilience \\

S8: Ouroboros &
Scarcity + Rule by Few + Hierarchical \newline Cluster 3 (Sustainability) &
Oscillatory civilization with cyclical collapse and regrowth \\

S9: Deus Ex Machina &
Non-scarcity + Rule by All + Distributed \newline Cluster 2 (Tech Escape) &
Aggressive technological expansion with instability risk \\

S10: Out of Eden &
Non-scarcity + Rule by All + Distributed \newline Cluster 1 (Parallel Life) &
Harmonized coexistence with nature and machines \\
\bottomrule
\end{tabular}
}

\smallskip
{\footnotesize $^{\dagger}$Cluster assignments follow \citet{Haqq-Misra2025-rt}; S3 shares the Earth~2024 technology cluster with S1 and S2 despite differing in global factors.\\
$^{\ddagger}$S7 is tagged non-scarcity in the original framework but parameterized with scarcity-level resource pressure to model post-collapse recovery (see text).}
\end{table*}

We assign parameter values based on the sociotechnical features of each scenario. The growth rate \(r\) is taken directly from \citet{Haqq-Misra2025-rt} and held fixed. We note that assuming technological capacity increases linearly each year via \(r\) is a simplification, especially given scenarios that imply discontinuities such as collapse points or radical transformation points; we retain it here for consistency with the source parameterization. The remaining parameters (initial resource stock \(R_{0}\), per-year depletion rate \(\delta\), collapse depth \(c_{f}\), recovery delay \(r_{d}\), recovery fraction \(r_{f}\), and existential hazard rate \(h\)) are derived from the narrative structure of each scenario and converted into quantitative values through structured reasoning. 

We assume that governance structure influences how technological civilizations respond to systemic stress and collapse. Highly centralized systems (rule by one) may retain moderate technological capacity through individual collapse events (higher $c_f$) due to concentrated infrastructure, and may recover relatively quickly (lower $r_d$) through top-down coordination. However, their rigidity and concentration of decision-making can make them more susceptible to repeated collapses when combined with resource scarcity or high external hazard exposure, consistent with the observation that increasing sociopolitical complexity yields diminishing returns, ultimately rendering centralized systems fragile \citep{Tainter1988-cc}. Oligarchic systems (rule by few) may experience deeper individual collapses (lower $c_f$) and slower recovery (higher $r_d$), reflecting fragmented coordination and weaker institutional memory. More distributed governance (rule by all) supports greater redundancy and flexibility, enabling a wider range of outcomes, from near-complete resilience in post-scarcity conditions to moderate collapse depth with relatively rapid recovery. This is consistent with the finding that polycentric governance systems mitigate the risk of institutional failure through overlapping jurisdictions and redundant decision-making structures \citep{Ostrom2010-pg}. To reflect these hypothesized patterns, we selected representative parameter value ranges for collapse depth \(c_{f}\), recovery delay \(r_{d}\), and recovery fraction \(r_{f}\) that align with the governance structure assumed in each scenario (Table~\ref{tab:governance}). These values are not based on empirical observation, but represent internally consistent assumptions designed to explore contrasting trajectories in our scenario-based simulations. 

One exception is the Restoration scenario (S7). While it is categorized as non-scarcity and governed by distributed institutions, it is conceptually framed as a rebuilding civilization emerging from an earlier collapse. To reflect this, we initialize the simulation with a reduced effective resource stock and a moderate depletion rate (parameters that would otherwise suggest scarcity), thereby modeling a society recovering from prior systemic failure. This allows us to represent S7 as a post-collapse recovery regime without forcing a collapse to occur within the simulation window.

The Ouroboros scenario (S8) is treated differently. In the original framework, S8 is described as a civilization that repeatedly undergoes expansion, collapse, and regrowth, with the year 3000 situated partway through an ongoing recovery phase. Rather than imposing externally timed or periodic collapse events, we model this behavior as an emergent dynamical regime. Specifically, S8 is assigned moderate technological growth, partial technological survival during collapse, and a relatively high recovery fraction of planetary resources. This combination allows repeated collapse--recovery cycles to arise endogenously from the interaction between resource depletion and rebuilding, and permits exploration of whether oscillatory behavior remains stable or dampens over long timescales.

Similarly, Deus Ex Machina (S9) is tagged as non-scarcity in the original framework but is parameterized with moderate resource pressure ($R_0 = 900$, $\delta = 1.3$) to reflect the high extraction demands of its aggressive technological expansion program, which drives resource depletion despite the scenario's nominally post-scarcity classification.

\begin{table}[htbp]
\centering
\caption{Mapping sociotechnical structure to parameter logic}
\label{tab:governance}
\begin{tabular}{llll}
\toprule
Governance Type & Collapse Depth $c_f$ & Recovery Delay $r_d$ & Recovery Fraction $r_f$ \\
\midrule
Rule by One & 0.5--0.7 & 40--60 & 0.10--0.20 \\
Rule by Few & 0.2--0.6 & 50--100 & 0.10--0.35 \\
Rule by All$^{*}$ & 0.3--1.0 & 0--70 & 0.20--1.0 \\
\bottomrule
\end{tabular}

\smallskip
{\footnotesize $^{*}$Post-scarcity idealizations (S3, S10) use $c_f = 1.0$, $r_d = 0$, $r_f = 1.0$, reflecting no-collapse conditions. S7 and S9 use parameters outside the typical governance range as narrative exceptions (see text).}
\end{table}

Scenarios also differ fundamentally in their assumptions about planetary resource conditions. Rather than define scarcity solely in terms of starting resources, we interpret it as a function of resource pressure (i.e., how quickly a civilization depletes its reserves). That is, systems with a large initial stock may still experience scarcity if it depletes resources at a high rate (e.g., S6). Conversely, a civilization with a modest stock may persist through repeated collapse--recovery cycles if post-collapse resource recovery partially replenishes its base (e.g., S8, S9), or may rebuild following an earlier systemic failure (e.g., S7).

\begin{table}[htbp]
\centering
\caption{Interpreting resource pressure from $R_0$ and $\delta$}
\label{tab:resources}
\begin{tabular}{llll}
\toprule
Resource Pressure & $R_0$ & $\delta$ & Scenarios \\
\midrule
High & $\leq 1000$ & 1.0--2.5 & S1, S2, S4, S7, S8, S9 \\
Moderate & 400--1600 & 0.5--1.0 & S6 \\
Low & $\geq 1200$ & 0.0--0.5 & S3, S5, S10 \\
\bottomrule
\end{tabular}
\end{table}

We implement Monte Carlo simulations to explore how uncertainty in model parameters shapes the long-term behavior of each scenario. Each scenario's baseline values (Table~\ref{tab:params}) are perturbed using probabilistic variation to capture plausible uncertainty (Table~\ref{tab:uncertainty}). Initial resource stock (\(R_{0}\)) and depletion rate (\(\delta\)) are sampled from normal distributions with modest standard deviations to represent uncertainty in ecological baselines and consumption patterns. Collapse depth (\(c_{f}\)) and recovery fraction (\(r_{f}\)) are sampled using triangular distributions centered on scenario-specific values, allowing outcomes to concentrate near expected levels while still capturing asymmetries in resilience and recovery capacity. Recovery delay (\(r_{d}\)) is sampled from a normal distribution with a standard deviation of $\pm 10$ years, reflecting variability in rebuilding pace across different societal structures.

The existential hazard rate (\(h\)) is held fixed for each scenario because it encodes an assumed background level of external risk (such as AI misalignment, interstellar conflict, or catastrophic environmental feedbacks) rather than internal system dynamics. Unlike other parameters whose variability arises from structural or operational uncertainty, \(h\) reflects a narrative-level judgment about a civilization's exposure to exogenous threats. Scenarios characterized by unstable geopolitical contexts, risky technological dependencies, or unresolved existential challenges (e.g., S1, S6, S7) are assigned higher hazard rates to reflect this vulnerability, while S5 carries a small residual risk due to high systemic complexity. In contrast, scenarios with societal stability, low-impact lifestyles, or resilient infrastructure (e.g., S3, S4, S8, S9, S10) are assigned zero or near-zero hazard rates. Keeping \(h\) fixed rather than sampling it probabilistically allows for clearer attribution of collapse outcomes to either endogenous fragility or exogenous shocks, rather than conflating the two.

\begin{table*}[htbp]
\centering
\caption{Final parameter values for each scenario}
\label{tab:params}
\begin{tabular}{lccccccc}
\toprule
Scenario & $r$ & $R_0$ & $\delta$ & $c_f$ & $r_d$ & $r_f$ & $h$ \\
\midrule
S1 & 0.0020 & 400 & 1.2 & 0.6 & 40 & 0.15 & 0.003 \\
S2 & 0.0026 & 700 & 1.5 & 0.3 & 70 & 0.10 & 0.001 \\
S3 & 0.0001 & 2400 & 0.0 & 1.0 & 0 & 1.0 & 0.0 \\
S4 & 0.0060 & 600 & 2.5 & 0.2 & 100 & 0.10 & 0.0 \\
S5 & 0.0003 & 2500 & 0.1 & 0.9 & 20 & 0.80 & 0.0005 \\
S6 & 0.0018 & 1600 & 1.0 & 0.4 & 50 & 0.25 & 0.005 \\
S7 & 0.0060 & 400 & 1.5 & 0.3 & 60 & 0.20 & 0.001 \\
S8 & 0.0015 & 800 & 1.3 & 0.6 & 70 & 0.35 & 0.0 \\
S9 & 0.0011 & 900 & 1.3 & 0.5 & 70 & 0.35 & 0.0 \\
S10 & 0.0002 & 2700 & 0.0 & 1.0 & 0 & 1.0 & 0.0 \\
\bottomrule
\end{tabular}
\end{table*}

\begin{table}[htbp]
\centering
\caption{Parameter uncertainty assumptions for Monte Carlo sampling}
\label{tab:uncertainty}
\begin{tabular}{lll}
\toprule
Parameter & Distribution & Uncertainty \\
\midrule
$R_0$ & Normal & $\pm 5\%$ of mean \\
$\delta$ & Normal & $\pm 5\%$ of mean \\
$c_f$ & Triangular & $\pm 5\%$ of mode \\
$r_d$ & Normal & $\pm 10$ years \\
$r_f$ & Triangular & $\pm 5\%$ of mode \\
\bottomrule
\end{tabular}
\end{table}

\FloatBarrier

\subsection{Metrics}\label{metrics}

We define four metrics to reflect different aspects of technospheric persistence and collapse dynamics across Monte Carlo simulations. These metrics are designed to capture the onset, frequency, and continuity of civilization-level activity, with a focus on their implications for detectability.

The first metric, \(T_{c}\), is the mean time to first collapse across replicates. It serves as a proxy for system fragility: scenarios with low values of \(T_{c}\) tend to break down early, while those with higher values demonstrate greater initial stability. When a simulation run does not experience collapse within the 1000-year window, the full timespan is used. The second metric, \(f_{nc}\), denotes the fraction of runs that avoid collapse entirely. This value reflects the overall robustness of a scenario and helps differentiate consistently stable futures from those vulnerable to systemic failure. We also measure \(D_{c}\), the average duty cycle, defined as the fraction of time a civilization remains active, that is, maintaining nonzero resource levels and thus potentially emitting technosignatures. This definition assumes that minimal resource availability is a prerequisite for detectable activity. Scenarios with high \(D_{c}\) maintain persistent activity, while those with low \(D_{c}\) may cycle through long dormant periods or experience early termination. Finally, \(N_{c}\), the mean number of collapse-recovery cycles, captures whether a scenario tends to exhibit singular collapse events or undergo repeated disruptions over time. Together, these metrics provide a compact but informative summary of each scenario's trajectory, enabling comparisons across sociotechnical regimes and offering insight into the reliability and intermittency of technosignature emission.

\section{Results}\label{results}

\subsection{Monte Carlo Outcomes and Temporal Profiles}\label{monte-carlo-outcomes-and-temporal-profiles}

\begin{figure*}[htbp]
\centering
\includegraphics[width=0.7\textwidth]{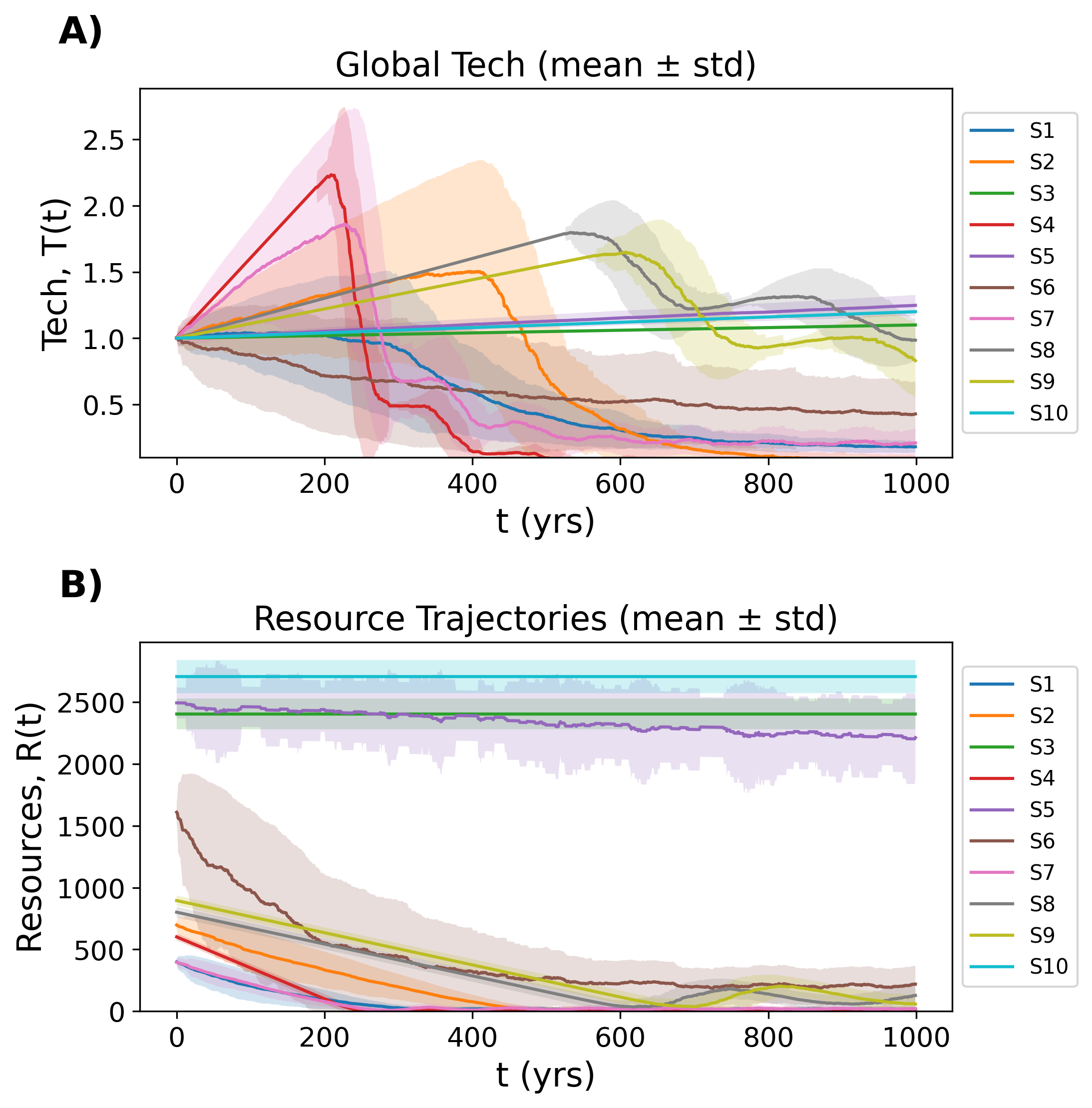}
\caption{Temporal dynamics of simulated civilizations across ten scenarios. (A) Mean trajectories of global technology \(T(t)\) with $\pm 1$ standard deviation shaded. (B) Mean resource levels \(R(t)\) with $\pm 1$ standard deviation shaded.}
\label{fig:trajectories}
\end{figure*}

We ran 200 independent simulations for each scenario, introducing probabilistic variation in model parameters as described in Section~\ref{model-description}. Figure~\ref{fig:trajectories} shows the resulting trajectories of global technology \(T(t)\) and available resources \(R(t)\), averaged across runs. These profiles reveal the distinct collapse and recovery dynamics that characterize each scenario. As expected, Golden Age (S3) and Out of Eden (S10) show uninterrupted growth and resource stability throughout the 1000-year window, consistent with their post-scarcity narratives and reflected in perfect duty cycles (\(\overline{D}_c = 1.0\)) and no collapse events (\(\overline{N}_c = 0\)). In stark contrast, scenarios such as Big Brother (S1) and Sword of Damocles (S6) suffer early collapses (\(\overline{T}_c < 210\)) and high collapse frequency (\(\overline{N}_c \approx 4.5\) for S6 and nearly 10 for S1), leading to sharp contractions in technological capacity and prolonged periods of inactivity. Deus Ex Machina (S9), although collapsing later (\(\overline{T}_c \approx 689\)), also experiences multiple interruptions and never recovers full continuity, with only modest duty cycle (\(\overline{D}_c \approx 0.91\)) and about 1.5 collapse events on average.

More complex dynamics emerge in intermediate cases. Transhumanism (S5) avoids collapse in roughly two-thirds of runs and achieves a near-continuous duty cycle (\(\overline{D}_c \approx 0.99\)), albeit with slower growth and occasional extinction. Living with the Land (S4), in contrast, undergoes frequent and deep collapses (\(\overline{N}_c \approx 6.6\)), producing the lowest sustained activity among all non-terminal scenarios (\(\overline{D}_c \approx 0.38\)) despite a steep initial growth phase. Restoration (S7), designed as a post-collapse recovery regime, displays consistent rebuilding behavior: early collapses (\(\overline{T}_c \approx 223\)), followed by approximately 7.5 recovery cycles and moderate continuity (\(\overline{D}_c \approx 0.57\)). Ouroboros (S8) exhibits a distinct pattern of low-frequency, self-stabilizing oscillations: it collapses only twice on average, yet maintains robust technological levels and a high duty cycle (\(\overline{D}_c \approx 0.87\)), reflecting a regime that cycles gracefully without exhausting resources or triggering runaway collapse.

\begin{figure*}[htbp]
\centering
\includegraphics[width=\textwidth]{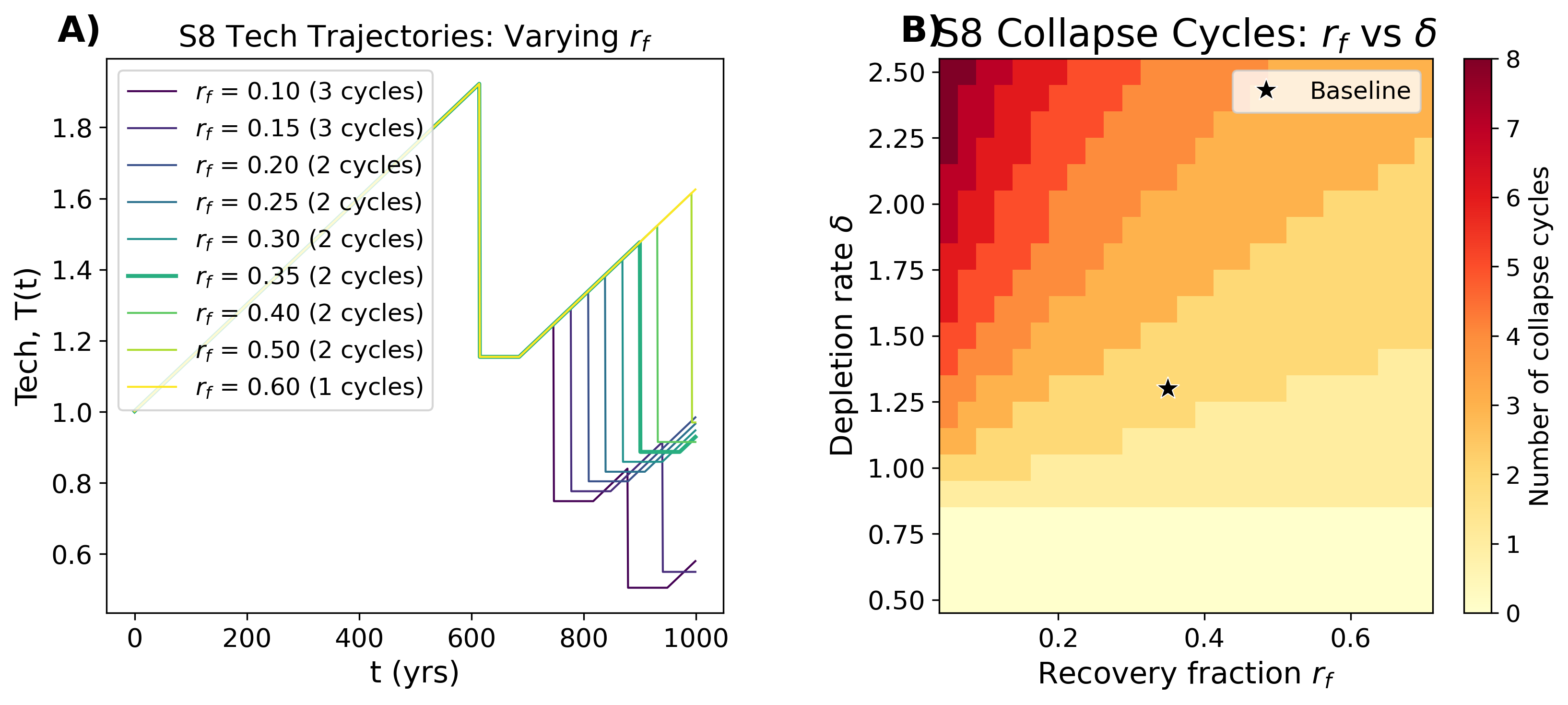}
\caption{Parameter sensitivity of S8 (Ouroboros) oscillatory dynamics. (A)~Deterministic technology trajectories $T(t)$ for varying recovery fraction $r_f$, with all other parameters held at baseline values. The number of collapse--recovery cycles within the 1000-year window is indicated in the legend. (B)~Number of collapse--recovery cycles as a function of recovery fraction $r_f$ and depletion rate $\delta$. The star marks the baseline S8 parameter combination.}
\label{fig:s8sweep}
\end{figure*}

Because the S8 narrative implies repeated collapse--recovery cycles, we ask whether those oscillations are always transient or can be sustained; we probe this with a sweep over $r_f$ and $\delta$ (Figure~\ref{fig:s8sweep}). Panel~A shows deterministic technology trajectories for varying $r_f$ while holding all other S8 parameters at their baseline values. For low recovery fractions ($r_f \leq 0.15$), the system undergoes rapid, repeated collapse--recovery cycles (3 cycles within the 1000-year window), as the diminished resource base after each recovery is quickly exhausted. At intermediate values ($0.20 \leq r_f \leq 0.50$), the system sustains two collapse--recovery cycles with longer growth phases between collapses. At higher recovery fractions ($r_f = 0.60$), the first collapse is delayed significantly, and only a single cycle occurs within the simulation window. Panel~B maps the number of collapse cycles across the $(r_f, \delta)$ parameter space, revealing a clear regime boundary: high depletion rates combined with low recovery fractions produce the most frequent oscillations (up to 8 cycles), while low depletion and high recovery fractions suppress collapse entirely. The baseline S8 parameters (marked with a star) sit near the boundary between 2- and 3-cycle regimes, suggesting that modest shifts in resource recovery capacity could qualitatively alter the civilization's long-term trajectory. These results confirm that S8's oscillatory behavior is not universally stable and it depends sensitively on the balance between resource depletion and post-collapse recovery, with some parameter combinations producing sustained cycling and others leading to dampened or terminal collapse.

\begin{figure*}[t]
\centering
\includegraphics[width=\textwidth]{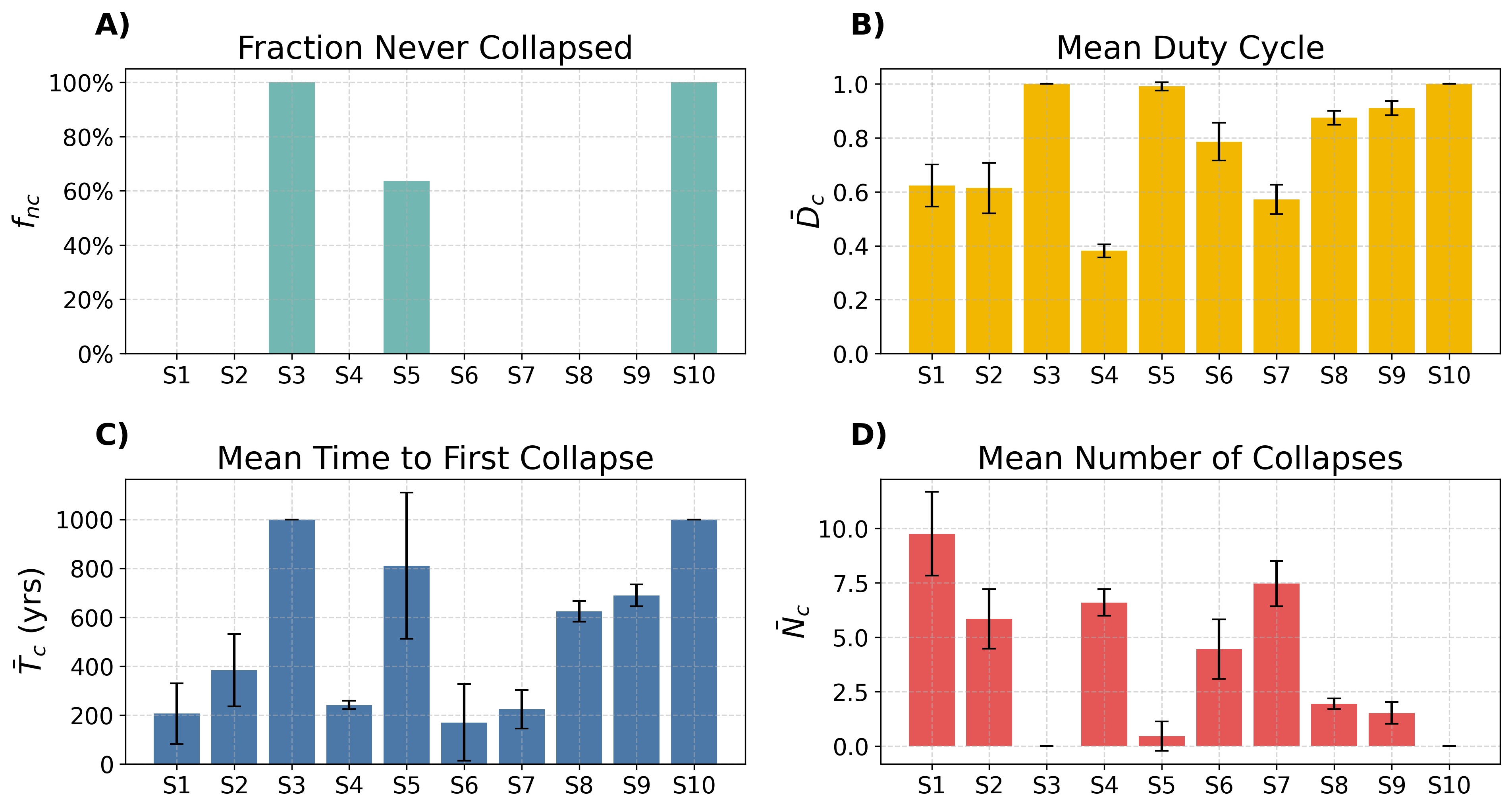}
\caption{Distributions of key collapse and recovery metrics across scenarios (200 runs each). (A) Fraction of simulations that never experienced collapse \(f_{nc}\). (B) Mean duty cycle \(\overline{D}_{c}\), defined as the fraction of time the technosphere remains active. (C) Mean time to first collapse \(\overline{T}_{c}\) with standard deviation bars. (D) Mean number of collapse events \(\overline{N}_{c}\).}
\label{fig:metrics}
\end{figure*}

Figure~\ref{fig:metrics} summarizes the outcomes of these simulations using the four metrics described above (\(\overline{T}_c\), \(f_{nc}\), \(\overline{D}_c\), \(\overline{N}_c\)). The scenarios divide into two distinct groups. S3, S5, and S10 survive the full 1000-year window in most or all runs (\(f_{nc} \geq 0.66\); Figure~\ref{fig:metrics}A), maintaining continuous or near-continuous technological activity (\(\overline{D}_c \geq 0.99\); Figure~\ref{fig:metrics}B). In contrast, the remaining seven scenarios collapse in every run, with S1 and S6 breaking down within the first 200 years on average (Figure~\ref{fig:metrics}C). Among the collapse-prone scenarios, the duty cycle spans a wide range---from 0.381 for S4 to 0.91 for S9---reflecting very different recovery capacities despite universal collapse. The most cyclical scenarios are S1 and S7, which undergo roughly 10 and 7.5 collapse--recovery cycles respectively (Figure~\ref{fig:metrics}D), while S8 cycles only about twice, consistent with its low-frequency oscillatory regime.

\section{Implications for Earth's Future}\label{earth-implications}

A key insight from our modeling is that the parameters governing recovery (the recovery fraction $r_f$ and the recovery delay $r_d$) are not fixed properties of a civilization but can be deliberately engineered. Investments in civilizational continuity infrastructure directly map onto these parameters: a higher $r_f$ corresponds to greater preservation of knowledge, technology, and resources through collapse events, while a lower $r_d$ corresponds to faster institutional and technological rebuilding. Together, improvements in these parameters increase the duty cycle $D_c$ and thus, increase the fraction of time civilization remains functional and productive.

The concept of "recovery kits" (curated repositories of essential knowledge designed to accelerate post-collapse rebuilding) represents one practical approach to raising $r_f$. \citet{Dartnell2014-tk} has explored in detail the minimum knowledge base required to reconstruct industrial civilization from scratch, spanning agriculture, materials science, medicine, power generation, and communication. Such recovery kits, whether physical or digital, function as a form of civilizational insurance: they ensure that the knowledge accumulated over centuries of development is not lost when institutional continuity is broken. Ongoing efforts in biological resource preservation serve as real-world examples of resilience infrastructure in action. The Svalbard Global Seed Vault, which now holds over one million seed samples from nearly every country on Earth \citep{Asdal2020-sv}, preserves agricultural biodiversity against civilizational disruption, ensuring that post-collapse societies retain access to the genetic resources needed to rebuild food systems. More recently, Colossal Biosciences has established a BioVault in Dubai dedicated to preserving broader biodiversity through cryopreservation of cellular material from threatened species \citep{Colossal2024-bv}. In the language of our model, these vaults serve to increase $r_f$ by preserving critical biological capital through collapse events.

In addition to interventions targeting $r_f$, it is also possible to concurrently engineer ways to decrease $r_d$, thereby helping societies rebuild faster by utilizing whatever fraction of resources can be salvaged. This vastly unexplored family of interventions concerns mechanisms of intergenerational communication and long-term memory transmission. If post-collapse societies cannot immediately (or even at all) make sense of the knowledge repositories and recovery kits that had been prudently engineered and purposely left behind to increase $r_f$, it will take them time to reverse-engineer the meaning necessary for actually using those repositories, assuming they were aware of their existence in the first place or accidentally came across them. This is related to the distinction between \emph{knowledge-that} (e.g., knowledge that this is a bike) and \emph{knowledge-how} (e.g., knowledge of how to ride a bike, in the sense of being able to ride it) \citep{Ren2012-kh}. Interventions that could decrease $r_d$ are related to the problem of communicating meanings to future humans that can ``endure for a very long time and, more importantly, be correctly decoded and interpreted'' \citep{Mazzucchelli2022-lf}. Although this problem has been mainly studied from the perspective of how to clearly communicate warnings about the dangers related to long-term storage sites of toxic waste \citep{Sandlos2019-gm} or nuclear waste \citep{Keating2023-ns}, the related learnings can be broadly applied to conceptualize such mechanisms. While the identification of promising combinations of exact mechanisms remains a fruitful research topic, such mechanisms appear to fall into two categories: indirect ones, such as education or surveillance, which aim at reinforcing the perpetuation of links between generations while working with social change; and direct ones, such as technical alterations of the environment or artifacts, which aim at immediately conveying information to future humans upon encountering such alterations or artifacts while remaining as much as possible independent of social change \citep{Calla2023-ww}.

Several other model inputs correspond to actionable levers for resilience policy. The depletion rate $\delta$ reflects the pace at which a civilization exhausts its resource base. Circular economy principles, renewable energy transitions, and efficiency improvements all serve to reduce $\delta$, extending the time before resource-driven collapse. This connects directly to recent work reframing civilizational growth in terms of thermodynamic efficiency and exploration-exploitation trade-offs \citep{Haqq-Misra2025-lum}. A civilization that pursues spatial expansion of its resource domain (exploration) effectively lowers its per-domain $\delta$, while one that intensifies extraction within a fixed domain (exploitation) increases it. The existential hazard rate $h$ can be reduced through investments in planetary defense (e.g., asteroid deflection programs such as NASA's Double Asteroid Redirection Test, which successfully demonstrated kinetic impact as a viable deflection strategy \citep{Daly2023-dart}), pandemic preparedness, AI safety research, and nuclear risk reduction, the last of which has been shown capable of triggering global famine affecting over five billion people through climate disruption alone \citep{Xia2022-nw}. \citet{Ord2020-tp} provides a systematic assessment of existential risks and argues that humanity's current exposure to catastrophic hazards is historically anomalous and reducible through deliberate policy. In our model, even small reductions in $h$ can significantly decrease the frequency of stochastic collapse events, particularly for scenarios like S1 and S6 where exogenous hazards are a primary driver of instability. The collapse survival fraction $c_f$ can be raised through distributed and redundant infrastructure. Decentralized energy grids, mesh communication networks, and dispersed governance structures all ensure that more technological capacity survives a systemic disruption. Finally, while the initial resource stock $R_0$ is treated as exogenous in our model, in practice it can be expanded through technologies that unlock previously inaccessible resources (e.g. asteroid mining, deep ocean extraction, nuclear fusion), effectively shifting a civilization into a lower resource-pressure regime.

Viewed through the duty-cycle framework, the question of civilizational resilience becomes quantifiable: what is the marginal improvement in $D_c$ from a given resilience investment? To address this, we perform a one-at-a-time sensitivity analysis across all seven collapse-prone scenarios, sweeping each human-controllable parameter from a low to a high plausible bound while holding all others at their scenario-specific baselines (Table~\ref{tab:sweep-ranges}).

\begin{figure*}[p]
\centering
\includegraphics[width=\textwidth,height=0.78\textheight,keepaspectratio]{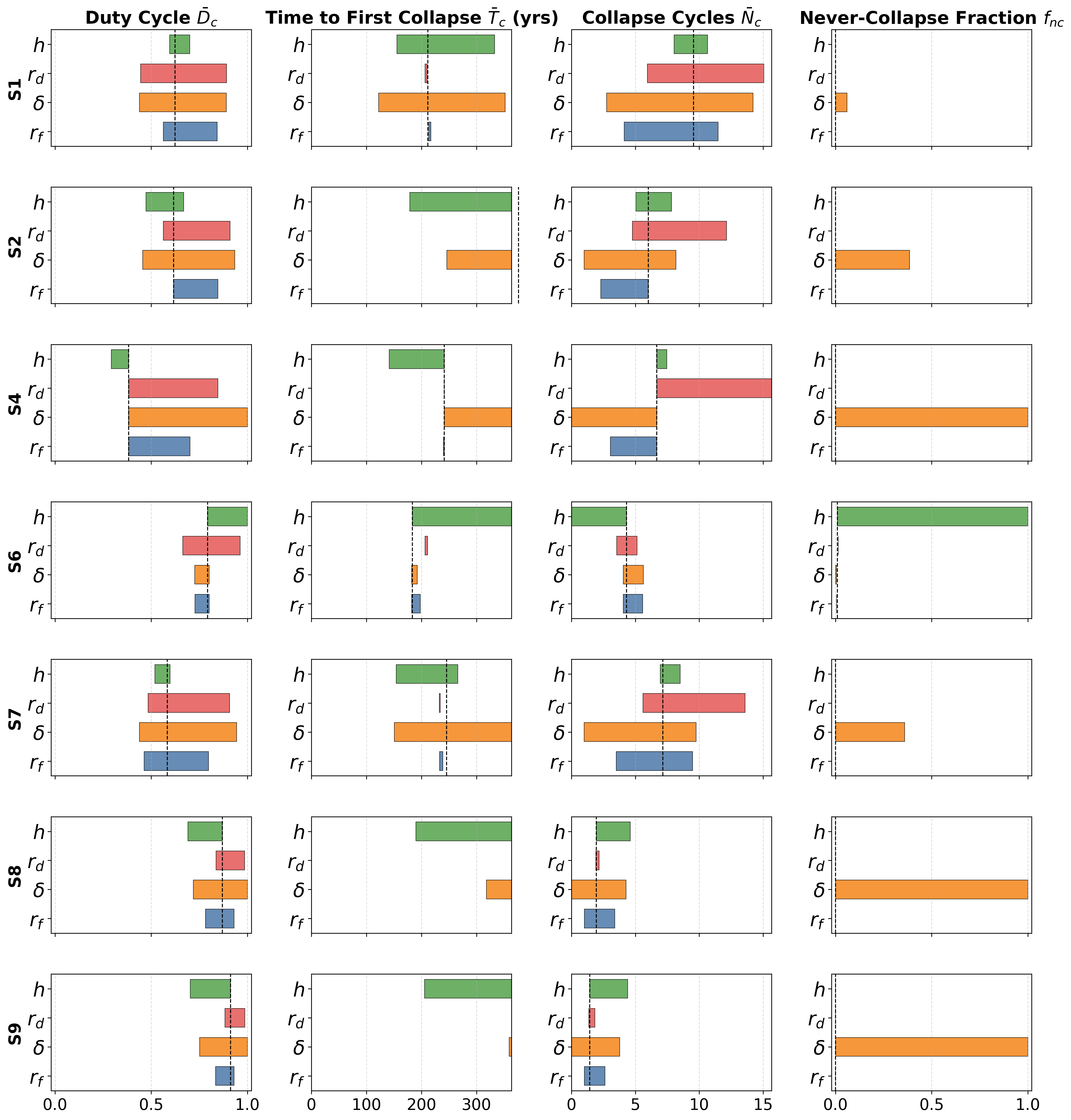}
\caption{Sensitivity analysis for all collapse-prone scenarios. Each row shows one scenario; columns show mean duty cycle ($\bar{D}_c$), mean time to first collapse ($\bar{T}_c$), mean number of collapse events ($\bar{N}_c$), and fraction of runs that never collapsed ($f_{nc}$). Bars span the outcome range when each parameter---recovery fraction ($r_f$), depletion rate ($\delta$), recovery delay ($r_d$), and existential hazard rate ($h$)---is swept from low to high bound; dashed lines mark baseline values. The collapse survival fraction ($c_f$) is omitted (negligible effect). Note that $h$ dominates in S6 where exogenous risk drives collapse, whereas $r_f$ and $\delta$ are the most impactful levers in resource-limited regimes.}
\label{fig:sensitivity-all}
\end{figure*}

Figure~\ref{fig:sensitivity-all} summarizes the results by showing, for each scenario, the swing in all four outcome metrics ($\bar{D}_c$, $\bar{T}_c$, $\bar{N}_c$, $f_{nc}$) when each parameter is moved from its low to its high bound. Across all scenarios, $r_f$ and $\delta$ consistently produce the largest swings, confirming them as the most impactful levers for civilizational resilience. The exception is S6, where the existential hazard rate $h$ dominates because exogenous risk, rather than resource depletion, is the primary driver of collapse.

\begin{figure*}[htbp]
\centering
\includegraphics[width=\textwidth,height=0.78\textheight,keepaspectratio]{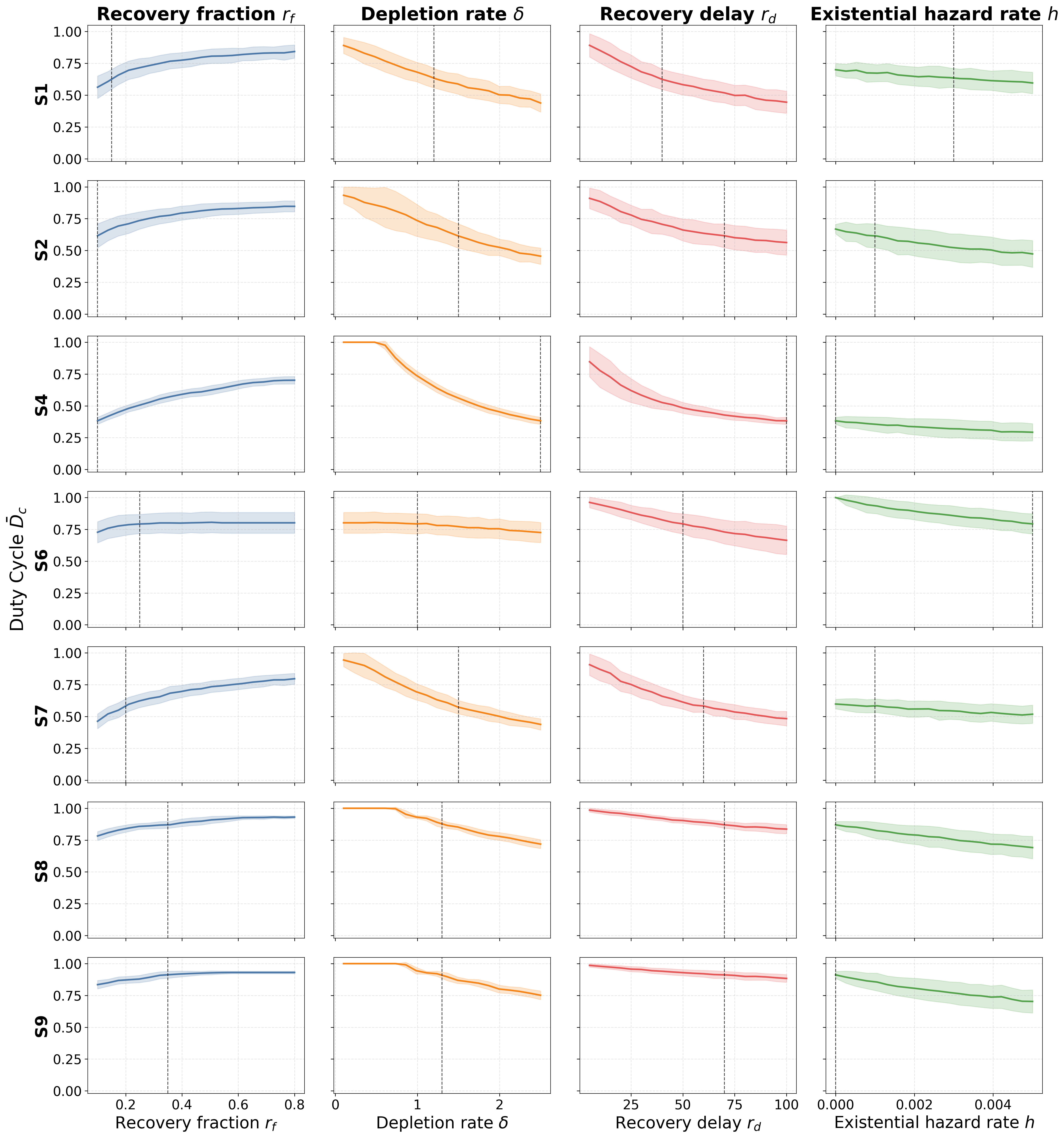}
\caption{Duty cycle ($\bar{D}_c$, fraction of time the technosphere remains active) as a continuous function of each human-controllable parameter for all collapse-prone scenarios (mean $\pm 1\sigma$ over 200 replicates). Each row shows one scenario; columns correspond to recovery fraction ($r_f$), depletion rate ($\delta$), recovery delay ($r_d$), and existential hazard rate ($h$). Dashed lines mark scenario-specific baseline values. The collapse survival fraction ($c_f$) is excluded (negligible effect).}
\label{fig:response-curves}
\end{figure*}

To examine the shape of these dependencies more closely, Figure~\ref{fig:response-curves} plots the duty cycle $\bar{D}_c$ as a continuous function of each parameter across all scenarios. Rather than a gradual, linear response, both $r_f$ and $\delta$ exhibit sharp nonlinear transitions: small changes near critical thresholds can shift the duty cycle from low to high values, while changes far from these thresholds have little effect. This implies that targeted resilience investments near the transition region could yield disproportionate returns in civilizational continuity. In contrast, $r_d$ and $h$ show more gradual responses across most scenarios. Together with the parameter sweep of S8 (Section~\ref{monte-carlo-outcomes-and-temporal-profiles}), these results demonstrate that the duty-cycle framework provides a practical metric for evaluating resilience investments.

\begin{table*}[htbp]
\centering
\caption{Parameter sweep ranges for sensitivity analysis. Low and high bounds are fixed across all scenarios; baseline values are scenario-specific. The collapse survival fraction ($c_f$, fraction of technology retained through a collapse event) is excluded because it has negligible effect on all outcome metrics.}
\label{tab:sweep-ranges}
\begin{tabular}{l c c c c c c c c c}
\toprule
 & \multicolumn{2}{c}{Sweep bounds} & \multicolumn{7}{c}{Baseline values} \\
\cmidrule(lr){2-3} \cmidrule(lr){4-10}
Parameter & Low & High & S1 & S2 & S4 & S6 & S7 & S8 & S9 \\
\midrule
$r_f$            & 0.10  & 0.80  & 0.15 & 0.10 & 0.10 & 0.25 & 0.20 & 0.35 & 0.35 \\
$\delta$         & 0.1   & 2.5   & 1.2  & 1.5  & 2.5  & 1.0  & 1.5  & 1.3  & 1.3  \\
$r_d$ (yr)       & 5     & 100   & 40   & 70   & 100  & 50   & 60   & 70   & 70   \\
$h$ (yr$^{-1}$)  & 0.0   & 0.005 & 0.003 & 0.001 & 0.0 & 0.005 & 0.001 & 0.0 & 0.0 \\
\bottomrule
\end{tabular}
\end{table*}

\section{Implications for the Search for Technosignatures}\label{technosignatures-implications}

\subsection{Reframing the Drake Equation's Longevity Term}\label{reframing-drake}

The Drake Equation \citep{Drake1965-wp} estimates the number of detectable civilizations in our galaxy as
\begin{equation}
N = R_*\, f_p\, n_e\, f_l\, f_i\, f_c\, L ,
\label{eq:drake}
\end{equation}
where \(R_{*}\) is the rate of star formation, \(f_{p}\) the fraction of those stars with planets, \(n_{e}\) the number of habitable planets per system, \(f_{l}\) the fraction of planets where life arises, \(f_{i}\) the fraction where intelligence evolves, \(f_{c}\) the fraction that develop detectable technology, and \(L\) the average duration such civilizations remain detectable. While the early terms are increasingly constrained by astrophysical observation, the final term \emph{L} carries outsized importance and remains deeply uncertain. Our simulations suggest that the traditional interpretation of $L$ as a continuous span of emission is unrealistic. To account for the intermittent nature of civilizational activity, we introduce an effective detectability duration,
\begin{equation}
L_{\mathrm{eff}} = D_c \, T_{\mathrm{span}} ,
\label{eq:leff}
\end{equation}
where \(D_{c}\) is the mean duty cycle and \(T_{\mathrm{span}}\) is the observation window (1000 years in our simulations). If $L$ is better understood as $L_{\mathrm{eff}}$, then models that assume continuous detectability may significantly overestimate $N$. A galaxy populated by civilizations with low duty cycles would remain largely silent, even if intelligent life is common. We note that a related but distinct concept, a ``reappearance factor'' for planets where civilizations can arise multiple times in succession \citep{Bhattacharya2011-qr}, which counts sequential independent civilizations over geological timescales, whereas our duty cycle refers to the active vs.\ inactive phases within a single civilization's lifespan.

Moreover, technology can outlast its creators, as the longevity of a technosignature may far exceed the lifespan of the civilization that produced it \citep{Wright2022-wh,Cirkovic2019-oq}. This further complicates the interpretation of $L$, which must account not only for active emission but also for the persistence of relic technosignatures, as we examine below.

These considerations help explain the Great Silence. Civilizations may exist in abundance, but if they are detectable only intermittently, the probability that their technosignatures overlap with our observational window is low \citep{Balbi2018-fp}. For two intermittent civilizations, the probability of mutual visibility is the product of their duty cycles, which can be very small. The simulation results raise a deeper question: should we expect civilizations to be uniform in character, or drawn from a diverse ensemble of trajectories? If most civilizations follow fragile, collapse-prone paths (e.g., S1, S2, S6), then long-duty-cycle civilizations would be statistical outliers, and the apparent absence of signals may reflect not the rarity of extraterrestrial technology but the dominance of intermittently silent technospheres. Our findings support this view: only S3 and S10 avoided collapse entirely, while most scenarios exhibited high collapse frequency and moderate or low duty cycles. These collapse-prone trajectories suggest that civilizations may not follow monotonic energy-growth trajectories, but instead oscillate or regress, challenging the static assumptions of the Kardashev scale \citep{Kardashev1964-gv}.

Recent work by \citet{Balbi2024-bg} provides further theoretical support for this view. Applying Lindy's law to technosignature longevities, they show that for energy-intensive technoemissions the longevity distribution follows a power law $\rho_L(L) \propto L^{-\alpha}$ with $\alpha > 1$, strongly disfavoring long-lived technosignatures. Under realistic assumptions ($\alpha \geq 2$), the first detected technosignature is predicted to be relatively short-lived ($L < 10^2$ years at 95\% confidence). If most detectable technosignatures are intrinsically short-lived, then the combination of Lindy-distributed longevities and low duty cycles (as observed in our collapse-prone scenarios) implies a doubly diminished probability of detection, offering a complementary explanation for the Great Silence.

\subsection{Technosignature Persistence and Observation Probabilities}\label{technosignature-persistence}

Eq. (\ref{eq:leff}) gives the detectability duration in terms of the mean duty cycle and time in the observation window. Some technosignatures may have a lifetime exactly equal to the the civilization's lifetime, but others may persist for some time after collapse. For technosignature $i$, the detectability duration is given as
\begin{equation}
    L_{i}=\{D_cT_{\text{span}} + \text{min}[T_i, (1-D_c)T_{\text{span}}]\}\delta_{i},
\end{equation}
where $T_i$ is the lifetime of technosignature $i$ and $\delta_{i}$ is a flag that is 1 if technosignature $i$ is present in the scenario and 0 otherwise.

\begin{figure*}[t]
\centering
\includegraphics[width=\textwidth]{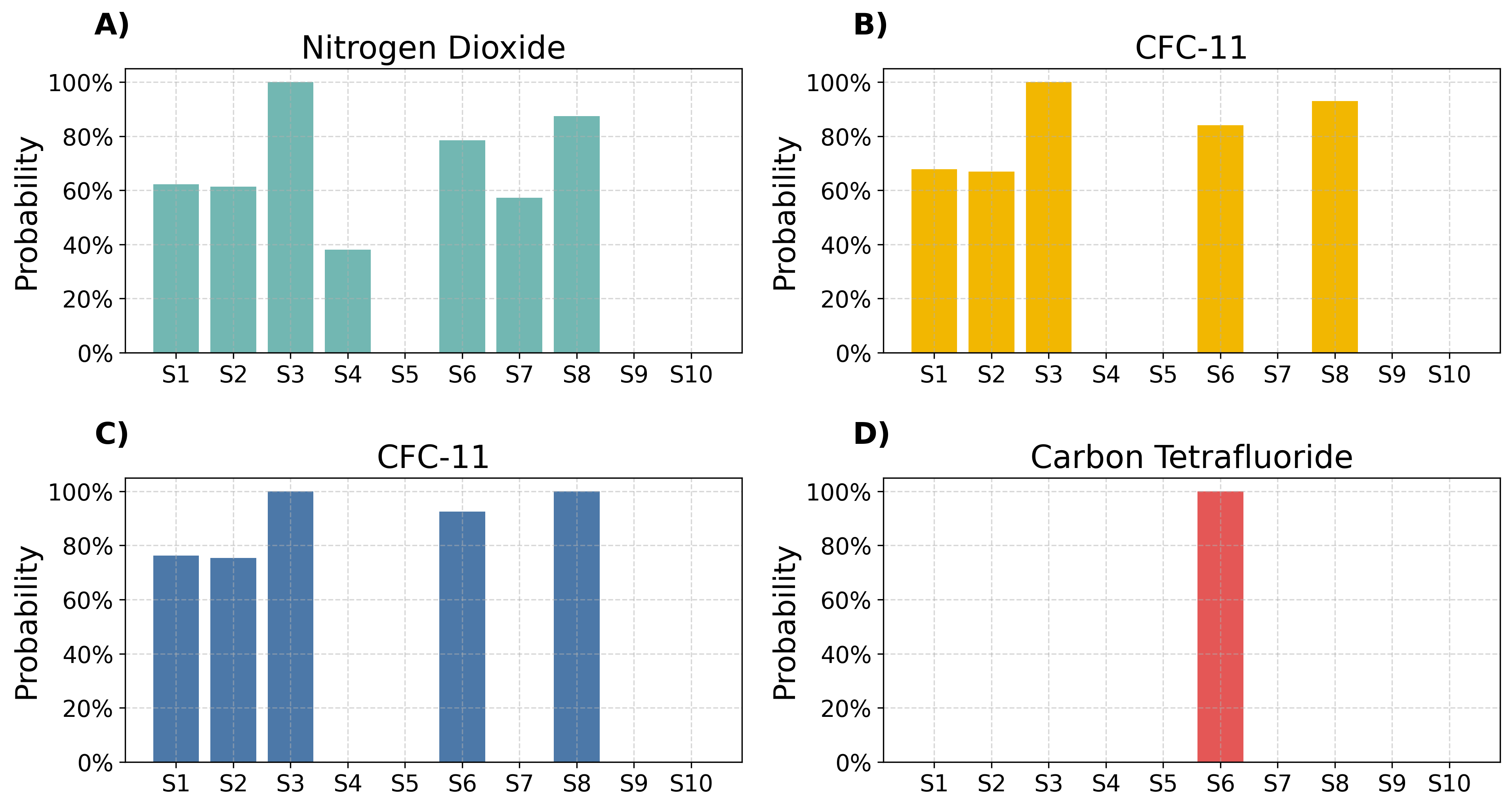}
\caption{The probability that a technosignature could be observed, taking into account the duty cycle for each scenario and the lifetime of each technosignature.}
\label{fig:ts_detect}
\end{figure*}

The probability that technosignature $i$ could be observed (assuming an appropriate observatory) is
\begin{equation}
    p_{i}=\frac{L_i}{T_{\text{span}}}\delta_{i} = \{D_c + \text{min}[\frac{T_i}{T_{\text{span}}}, 1-D_c]\}\delta_{i},
\end{equation}
These probabilities are shown in Figure \ref{fig:ts_detect} for four example technosignatures, with a 1000-year time span. These probabilities should be interpreted as the probability of finding a given technosignature for remote observations of Earth in a given scenario, assuming the use of a capable observatory and sufficient observation time. The values of $\delta_{i}$ were determined from \citet[][Table 7]{Haqq-Misra2025-rt}. Nitrogen dioxide (NO$_2$) has a short lifetime of hours to days, so the probability is about equal to the duty cycle for scenarios in which nitrogen dioxide is present. The lifetime of CFC-11 is 55\,yr, and the lifetime of CFC-12 is 140\,yr, which could be observed in five of the scenarios. CFC-12 has higher observation probabilities due to its longer lifetime. Carbon tetrafluoride (CF$_4$) has a lifetime of 1000\,yr or longer, observable in one of the scenarios. These are examples of atmospheric technosignatures that are present in the scenario set and could be observable by mission concepts under study such as the Habitable Worlds Observatory and the Large Interferometer for Exoplanets \citep{haqq2025projections}.

The $L_{\mathrm{eff}}$ formulation assumes that detectability ends when activity ceases. As noted above, this need not be the case \citep{Balbi2021-bc}. The persistence of different technosignature types varies enormously. Orbital debris and artificial satellites can last from decades to as long as 10,000 to 100,000 years. Satellites in low Earth orbit typically deorbit within decades due to atmospheric drag, while those in geosynchronous orbit, facing minimal resistance, may persist for much longer. Without maintenance, a dense band of satellites in the so-called Clarke belt would slowly degrade through collisions and orbital perturbations over tens of thousands of years, eventually becoming undetectable \citep{Socas-Navarro2018-kg}. Megastructures such as Dyson spheres, space mirrors, or stellar engines may persist longer still, though their survival depends on structural design and orbital stability. Rigid constructs without active station-keeping may collapse or drift within a few centuries to several thousand years \citep{Wright2020-be}, however, more modular configurations, like swarms of solar collectors, could remain in stable orbits for far longer timescales \citep{Smith2022-ds}.

Atmospheric technosignatures provide the shortest windows of detectability. On Earth, CFC-11 and CFC-12 have atmospheric lifetimes of about 55 and 140 years, respectively, before being broken down by photolysis or other chemical processes \citep{NOAA-GML-lz}. Other compounds are more resilient: tetrafluoromethane (CF$_4$) persists for \textasciitilde50,000 years, hexafluoroethane (C$_2$F$_6$) for \textasciitilde10,000 years, and octafluoropropane (C$_3$F$_8$) for \textasciitilde2,600 years \citep{Trudinger2016-ky}. Longer-lived isotopic signatures are conceivable, but their remote detection remains technically challenging. Electromagnetic technosignatures such as radio or laser transmissions propagate indefinitely but are only practically detectable for finite periods. Focused, high-power signals might remain observable for a few hundred light-years over spans of centuries to millennia, depending on the strength of the signal and the sensitivity of receivers \citep{Sheikh2025-td}.

These persistence times imply that technosignature lifetimes can range from negligible (radio bursts) to dominant (orbital infrastructure). For short-lived civilizations with brief ON phases, persistent technosignatures could substantially extend the effective detectability window beyond what $L_{\mathrm{eff}}$ alone would suggest. In contrast, for continuously active civilizations, persistence contributes relatively little. If most detectable technosignatures are short-lived (e.g., radio leakage), then the duty cycle remains the key parameter, and our results suggest low duty cycles make detection unlikely. But if many technosignatures persist for millennia beyond collapse, then the silence we observe becomes harder to attribute solely to civilizational fragility; it would require that either technosignatures are intrinsically ephemeral, or civilizations rarely produce persistent artifacts.

In summary, our results suggest that if detectability is often sporadic, then long periods without contact may be the expected baseline. Intermittency may be typical, and recognizing that could refine our expectations and expand the scope of SETI. We also note that some persistent technologies, particularly autonomous systems such as Bracewell probes \citep{Bracewell1960-bp} or orbital maintenance infrastructure, may not decay passively but instead cycle through their own active and dormant states, creating layered duty-cycle structures that merit further investigation.

\section{Limitations}\label{limitations}

Several limitations warrant discussion. Our modeling framework is grounded in Earth-originating scenario typologies, drawing from narrative futures methodologies informed by contemporary geopolitical, ecological, and technological trends. These projections embed assumptions about governance structures, resource constraints, and recovery dynamics that may not extend to non-terrestrial civilizations or to those that do not follow Earth-like developmental pathways. While this grounding enables empirical plausibility, it limits extrapolation beyond our biosphere's trajectory \citep{Rubin2001-sk,Denning2011-fj}. Our simulations assume bounded planetary systems and do not account for galactic colonization, interstellar migration, or modes of expansion that could alter the spatial and temporal visibility of civilizations \citep{Walters1980-ev,Papagiannis1983-xv}. We also exclude the effects of cultural convergence, information saturation, or technological transcendence (e.g., the emergence of post-biological intelligence), all of which could suppress or qualitatively change technosignature output. Our operational definition of detectability (based on internal sociotechnical continuity) does not differentiate among specific technosignature classes (e.g., radio emissions, waste heat, megastructures), nor does it address detection thresholds or instrumental sensitivity.

Additionally, our assumption that technological capacity increases linearly each year according to a fixed growth rate $r$ is a significant simplification. The many discontinuities implied by the scenarios (not only collapse points but also radical transformation points such as the emergence of artificial general intelligence or breakthroughs in energy harvesting) suggest that a linear growth model may not adequately capture the dynamics of technological change across all regimes. Moreover, the available resource stock $R_0$ is treated as an exogenous parameter, but in practice it is contingent on the technological capacity to identify and utilize resources. For example, a civilization that develops asteroid mining or nuclear fusion effectively increases its accessible resource base, fundamentally altering the growth-depletion dynamics. As discussed in Section~\ref{earth-implications}, \citet{Haqq-Misra2025-lum} reframe the Kardashev scale as a luminosity limit constrained by thermodynamic efficiency (exergy) and explore exploration-exploitation trade-offs whose distinction our linear growth model does not capture. Future work could incorporate exergy constraints and non-linear growth functions informed by their luminosity-limit model, as well as feedback between technological capacity and accessible resource stocks, to produce more physically grounded projections of civilizational trajectories.

\section{Conclusions}

We have developed a hybrid deterministic-stochastic model to simulate collapse-recovery dynamics across ten plausible futures for Earth-originating civilization, and used these simulations to quantify the duty cycles of technological civilizations. Our principal findings are as follows.

First, collapse-recovery dynamics produce a wide range of duty cycles across plausible civilizational futures, with $D_c$ spanning from approximately 0.38 to 1.00 across the ten scenarios, with outcomes shaped by the interplay of governance structure, resource pressure, and hazard exposure.

Second, the duty-cycle framework reveals that several model parameters correspond to actionable levers for civilizational resilience (Section~\ref{earth-implications}). Our sensitivity analysis demonstrates that $r_f$ and $\delta$ consistently emerge as the most impactful parameters across all collapse-prone scenarios, and that modest shifts in these levers can qualitatively alter a civilization's long-term trajectory, suggesting that targeted resilience investments could yield disproportionate returns in civilizational continuity.

Third, the effective detectability duration $L_{\mathrm{eff}} = D_c \, T_{\mathrm{span}}$ provides a more realistic replacement for the Drake Equation's static longevity term $L$. Combined with power-law longevity distributions that disfavor long-lived technosignatures \citep{Balbi2024-bg}, low duty cycles imply a doubly diminished probability of detection, helping explain the observed absence of extraterrestrial signals.

Several avenues for future work emerge from this study. Incorporating exergy constraints and non-linear growth functions informed by thermodynamic limits to growth \citep{Haqq-Misra2025-lum} could yield more physically grounded simulations. Introducing feedback between technological capacity and accessible resource stocks would address a key simplification of the current model. Finally, extending the framework to account for per-technosignature observation probabilities and for the possibility that autonomous remnant technologies may cycle through their own active and dormant states would add further realism to the detectability analysis.

More broadly, our results help reconcile seemingly contradictory perspectives on civilizational longevity. One view posits that technological civilizations are fundamentally fragile and prone to self-destruction soon after emergence \citep{Vinn2024-pi}. Another suggests that once civilizations surpass key thresholds of resilience and coordination, they may persist indefinitely, steadily accumulating across cosmic time \citep{Grinspoon2009-cg}. Both patterns emerge naturally in our simulations---not from differences in exogenous shocks, but from internal sociotechnical structure. The long-term fate of a civilization, it appears, is less a matter of luck than of design.
\section*{Code Availability}
All code used to run the simulations and generate the figures in this paper is publicly available at \url{https://github.com/celiablanco/technocycles}.

\section*{Acknowledgments}
C.B.\ acknowledges financial support from Ram\'{o}n y Cajal grant RYC2023-045781-I funded by the Spanish Agencia Estatal de Investigaci\'{o}n MICIU/AEI/10.13039/501100011033 and by ``ESF Investing in your future'', and from the RyC-MAX grant 20255AT022 funded by the Spanish National Research Council.

\bibliographystyle{plainnat}
\bibliography{references}

@ARTICLE{Ren2012-kh,
  title     = "The distinction between knowledge-that and knowledge-how",
  author    = "Ren, Huiming",
  journal   = "Philosophia",
  volume    =  40,
  number    =  4,
  pages     = "857--875",
  year      =  2012,
  doi = "10.1007/s11406-012-9361-x"
}

@ARTICLE{Mazzucchelli2022-lf,
  title     = "How to remember a place to forget? {The} semiotic design of deep
               geological nuclear repositories, from long-term communication to
               memory transmission",
  author    = "Mazzucchelli, Francesco and Novello Paglianti, Nanta",
  journal   = "Linguistic Frontiers",
  volume    =  5,
  number    =  3,
  pages     = "22--36",
  year      =  2022,
  doi = "10.2478/lf-2022-0026"
}

@ARTICLE{Sandlos2019-gm,
  title     = "There is a Monster Under the Ground: Commemorating the History
               of Arsenic Contamination at {Giant Mine} as a Warning to Future
               Generations",
  author    = "Sandlos, John and Keeling, Arn and Beckett, Caitlynn and Nicol, Rosanna",
  journal   = "Papers in Canadian History and Environment",
  number    =  3,
  pages     = "1--55",
  month     =  oct,
  year      =  2019,
  doi = "10.25071/10315/36516"
}

@ARTICLE{Keating2023-ns,
  title     = "Nuclear memory: Archival, aesthetic, speculative",
  author    = "Keating, Thomas P and Storm, Anna",
  journal   = "Progress in Environmental Geography",
  volume    =  2,
  number    = "1-2",
  pages     = "97--117",
  year      =  2023,
  doi = "10.1177/27539687231174242"
}

@ARTICLE{Calla2023-ww,
  title     = "Confronting the Uncertainties Associated with Long-Time Scales:
               Analysis of the Modes of Preservation of Memory of Radioactive
               Waste Burial Sites",
  author    = "Calla, S and Guinchard, C and Moine, A and Novello-Paglianti, N
               and Nuninger, L and Ogorzelec-Guinchard, L",
  journal   = "Worldwide Waste: Journal of Interdisciplinary Studies",
  volume    =  6,
  number    =  1,
  pages     = "1--12",
  year      =  2023,
  doi = "10.5334/wwwj.75"
}

@article{mullan2019population,
  title={Population growth, energy use, and the implications for the search for extraterrestrial intelligence},
  author={Mullan, Brendan and Haqq-Misra, Jacob},
  journal={Futures},
  volume={106},
  pages={4--17},
  year={2019},
  publisher={Elsevier}
}

@article{haqq2025projections,
  title={Projections of Earth’s Technosphere: Strategies for Observing Technosignatures on Terrestrial Exoplanets},
  author={Haqq-Misra, Jacob and Kopparapu, Ravi K and Profitiliotis, George},
  journal={The Astrophysical Journal Letters},
  volume={995},
  number={1},
  pages={L22},
  year={2025},
  publisher={The American Astronomical Society}
}

@ARTICLE{Cirkovic2019-oq,
  title         = "Persistence of Technosignatures: A Comment on Lingam and
                   Loeb",
  author        = "{\'C}irkovi{\'c}, M. M. and Vukoti{\'c}, B. and
                   Stojanovi{\'c}, M.",
  month         =  may,
  year          =  2019,
  copyright     = "http://arxiv.org/licenses/nonexclusive-distrib/1.0/",
  archivePrefix = "arXiv",
  primaryClass  = "physics.pop-ph",
  eprint        = "1905.03146",
  doi = "10.48550/arXiv.1905.03146"
}

@ARTICLE{Vinn2024-pi,
  title     = "Potential incompatibility of inherited behavior patterns with
               civilization: Implications for Fermi paradox",
  author    = "Vinn, O.",
  journal   = "Sci. Prog.",
  publisher = "SAGE Publications",
  volume    =  107,
  number    =  3,
  pages     = "368504241272491",
  month     =  jul,
  year      =  2024,
  doi = "10.1177/00368504241272491"
}

@ARTICLE{Kardashev1964-gv,
  title   = "Transmission of information by extraterrestrial civilizations",
  author  = "Kardashev, N.",
  journal = "SvA",
  volume  =  8,
  pages   = "217",
  month   =  oct,
  year    =  1964
}

@ARTICLE{Haqq-Misra2025-rt,
  title     = "Projections of Earth's technosphere: Scenario modeling,
               worldbuilding, and overview of remotely detectable
               technosignatures",
  author    = "Haqq-Misra, J. and Profitiliotis, G. and Kopparapu, R.",
  journal   = "Technol. Forecast. Soc. Change",
  publisher = "Elsevier BV",
  volume    =  218,
  number    =  124194,
  pages     = "124194",
  month     =  sep,
  year      =  2025,
  copyright = "http://creativecommons.org/licenses/by/4.0/",
  doi = "10.1016/j.techfore.2025.124194"
}

@ARTICLE{Walters1980-ev,
  title     = "Interstellar colonization: A new parameter for the Drake
               equation?",
  author    = "Walters, C. and Hoover, R. A. and Kotra, R. K.",
  journal   = "Icarus",
  publisher = "Elsevier BV",
  volume    =  41,
  number    =  2,
  pages     = "193--197",
  month     =  feb,
  year      =  1980,
  doi = "10.1016/0019-1035(80)90003-2"
}

@ARTICLE{Haqq-Misra2009-jb,
  title    = "The Sustainability Solution To The Fermi Paradox",
  author   = "Haqq-Misra, J. D. and Baum, S. D.",
  journal  = "J. Br. Interplanet. Soc.",
  volume   =  62,
  number   =  2,
  pages    = "47--51",
  year     =  2009
}

@ARTICLE{Papagiannis1983-xv,
  title     = "The importance of exploring the asteroid belt",
  author    = "Papagiannis, M. D.",
  journal   = "Acta Astronaut.",
  publisher = "Elsevier BV",
  volume    =  10,
  number    =  10,
  pages     = "709--712",
  month     =  oct,
  year      =  1983,
  doi = "10.1016/0094-5765(83)90071-1"
}

@INCOLLECTION{Drake1965-wp,
  title     = "The radio search for intelligent extraterrestrial life",
  booktitle = "Current Aspects of Exobiology",
  author    = "Drake, F. D.",
  publisher = "Elsevier",
  pages     = "323--345",
  year      =  1965,
  doi = "10.1016/b978-1-4832-0047-7.50015-0"
}

@ARTICLE{Socas-Navarro2018-kg,
  title     = "Possible photometric signatures of moderately advanced
               civilizations: The Clarke exobelt",
  author    = "Socas-Navarro, H.",
  journal   = "Astrophys. J.",
  publisher = "American Astronomical Society",
  volume    =  855,
  number    =  2,
  pages     = "110",
  month     =  mar,
  year      =  2018,
  copyright = "http://iopscience.iop.org/page/copyright",
  doi = "10.3847/1538-4357/aaae66"
}

@ARTICLE{Bhattacharya2011-qr,
  title   = "Some computations on the Drake equation to encapsulate the
             probable number of broadcasting civilizations",
  author  = "Bhattacharya, A. B. and Sarkar, A. and Pandit, J.",
  journal = "International Journal of Applied Engineering Research",
  volume  =  2,
  number  =  2,
  pages   = "590--594",
  year    =  2011
}

@ARTICLE{Wong2022-hv,
  title     = "Asymptotic burnout and homeostatic awakening: a possible
               solution to the Fermi paradox?",
  author    = "Wong, M. L. and Bartlett, S.",
  journal   = "J. R. Soc. Interface",
  publisher = "The Royal Society",
  volume    =  19,
  number    =  190,
  pages     = "20220029",
  month     =  may,
  year      =  2022,
  doi = "10.1098/rsif.2022.0029"
}

@ARTICLE{Balbi2018-fp,
  title     = "The impact of the temporal distribution of communicating
               civilizations on their detectability",
  author    = "Balbi, A.",
  journal   = "Astrobiology",
  publisher = "Mary Ann Liebert Inc",
  volume    =  18,
  number    =  1,
  pages     = "54--58",
  month     =  jan,
  year      =  2018,
  doi = "10.1089/ast.2017.1652"
}

@ARTICLE{Gray2020-cn,
  title     = "Intermittent signals and planetary days in {SETI}",
  author    = "Gray, R. H.",
  journal   = "Int. J. Astrobiology",
  publisher = "Cambridge University Press (CUP)",
  volume    =  19,
  number    =  4,
  pages     = "299--307",
  month     =  aug,
  year      =  2020,
  doi = "10.1017/s1473550420000038"
}

@BOOK{Grinspoon2009-cg,
  title     = "Lonely planets",
  author    = "Grinspoon, D.",
  publisher = "HarperCollins eBooks",
  month     =  mar,
  year      =  2009
}

@BOOK{Webb2015-vi,
  title     = "If the Universe Is Teeming with Aliens ... {WHERE} {IS}
               {EVERYBODY}? Seventy-Five Solutions to the Fermi Paradox
               and the Problem of Extraterrestrial Life",
  author    = "Webb, S.",
  publisher = "Springer International Publishing",
  year      =  2015,
  edition   = "2nd",
  doi = "10.1007/978-3-319-13236-5"
}

@ARTICLE{Denning2011-fj,
  title     = "Being technological",
  author    = "Denning, K.",
  journal   = "Acta Astronaut.",
  publisher = "Elsevier BV",
  volume    =  68,
  number    = "3-4",
  pages     = "372--380",
  month     =  feb,
  year      =  2011,
  doi = "10.1016/j.actaastro.2010.02.027"
}

@INPROCEEDINGS{Rubin2001-sk,
  title      = "{L} factor: hope and fear in the search for extraterrestrial
                intelligence",
  booktitle  = "The Search for Extraterrestrial Intelligence ({SETI}) in the
                Optical Spectrum {III}",
  author     = "Rubin, C. T.",
  editor     = "Kingsley, S. A. and Bhathal, R.",
  publisher  = "SPIE",
  volume     =  4273,
  pages      = "230--239",
  month      =  aug,
  year       =  2001,
  conference = "Photonics West 2001 - LASE",
  location   = "San Jose, CA",
  doi = "10.1117/12.435379"
}

@ARTICLE{Shermer2002-oi,
  title     = "Why {ET} Hasn't Called",
  author    = "Shermer, M.",
  journal   = "Sci. Am.",
  publisher = "Springer Science and Business Media LLC",
  volume    =  287,
  number    =  2,
  pages     = "33--33",
  month     =  aug,
  year      =  2002,
  doi = "10.1038/scientificamerican0802-33"
}

@ARTICLE{Wright2022-wh,
  title     = "The case for technosignatures: Why they may be abundant,
               long-lived, highly detectable, and unambiguous",
  author    = "Wright, J. T. and Haqq-Misra, J. and Frank, A. and
               Kopparapu, R. and Lingam, M. and Sheikh, S. Z.",
  journal   = "Astrophys. J. Lett.",
  publisher = "American Astronomical Society",
  volume    =  927,
  number    =  2,
  pages     = "L30",
  month     =  mar,
  year      =  2022,
  copyright = "http://creativecommons.org/licenses/by/4.0/",
  doi = "10.3847/2041-8213/ac5824"
}

@ARTICLE{Sheikh2025-td,
  title     = "Earth detecting Earth: At what distance could Earth's
               constellation of technosignatures be detected with present-day
               technology?",
  author    = "Sheikh, S. Z. and Huston, M. J. and Fan, P. and Wright,
               J. T. and Beatty, T. and Martini, C. and Kopparapu,
               R. and Frank, A.",
  journal   = "Astron. J.",
  publisher = "American Astronomical Society",
  volume    =  169,
  number    =  2,
  pages     = "118",
  month     =  feb,
  year      =  2025,
  copyright = "https://creativecommons.org/licenses/by/4.0/",
  doi = "10.3847/1538-3881/ada3c7"
}

@ARTICLE{Trudinger2016-ky,
  title     = "Atmospheric abundance and global emissions of perfluorocarbons
               {CF$_{4}$}, {C$_{2}$F$_{6}$} and {C$_{3}$F$_{8}$} since 1800
               inferred from ice core, firn, air archive and in situ
               measurements",
  author    = "Trudinger, C. M. and Fraser, P. J. and Etheridge, D. M. and others",
  journal   = "Atmos. Chem. Phys.",
  publisher = "Copernicus GmbH",
  volume    =  16,
  number    =  18,
  pages     = "11733--11754",
  month     =  sep,
  year      =  2016,
  copyright = "https://creativecommons.org/licenses/by/3.0/",
  doi = "10.5194/acp-16-11733-2016"
}

@ARTICLE{Wright2020-be,
  title     = "Dyson spheres",
  author    = "Wright, J. T.",
  journal   = "Serbian Astron. J.",
  publisher = "National Library of Serbia",
  number    =  200,
  pages     = "1--18",
  year      =  2020,
  doi = "10.2298/saj2000001w"
}

@ARTICLE{Wesson1990-xi,
  title   = "Cosmology, extraterrestrial intelligence, and a resolution of the
             Fermi-hart paradox",
  author  = "Wesson, P.",
  journal = "Quarterly journal of the royal astronomical society",
  volume  =  31,
  pages   = "161--170",
  month   =  jun,
  year    =  1990
}

@ARTICLE{Balbi2021-bc,
  title     = "Longevity is the key factor in the search for technosignatures",
  author    = "Balbi, A. and {\'C}irkovi{\'c}, M. M.",
  journal   = "Astron. J.",
  publisher = "American Astronomical Society",
  volume    =  161,
  number    =  5,
  pages     = "222",
  month     =  may,
  year      =  2021,
  doi = "10.3847/1538-3881/abec48"
}

@ARTICLE{Balbi2024-bg,
  title     = "Technosignature longevity and Lindy's law",
  author    = "Balbi, A. and Grimaldi, C.",
  journal   = "Astron. J.",
  publisher = "American Astronomical Society",
  volume    =  167,
  number    =  3,
  pages     = "119",
  month     =  mar,
  year      =  2024,
  doi = "10.3847/1538-3881/ad217d"
}

@ARTICLE{Haqq-Misra2025-lum,
  title     = "Projections of Earth's technosphere: Luminosity and mass as
               limits to growth",
  author    = "Haqq-Misra, J. and Vidal, C. and Profitiliotis, G.",
  journal   = "Acta Astronaut.",
  publisher = "Elsevier BV",
  volume    =  229,
  pages     = "831--838",
  year      =  2025,
  doi = "10.1016/j.actaastro.2025.01.048"
}

@BOOK{Dartnell2014-tk,
  title     = "The Knowledge: How to Rebuild Our World from Scratch",
  author    = "Dartnell, L.",
  publisher = "Penguin Press",
  year      =  2014
}

@ARTICLE{Asdal2020-sv,
  title     = "The {Svalbard} {Global} {Seed} {Vault}: 10 Years---1 Million
               Samples",
  author    = "Asdal, {\AA}. and Guarino, L.",
  journal   = "Biopreserv. Biobank.",
  volume    =  18,
  number    =  5,
  pages     = "391--392",
  year      =  2020,
  doi = "10.1089/bio.2018.0025"
}

@BOOK{Ord2020-tp,
  title     = "The Precipice: Existential Risk and the Future of Humanity",
  author    = "Ord, T.",
  publisher = "Hachette Books",
  year      =  2020
}

@BOOK{Tainter1988-cc,
  title     = "The Collapse of Complex Societies",
  author    = "Tainter, J. A.",
  publisher = "Cambridge University Press",
  year      =  1988
}

@ARTICLE{Ostrom2010-pg,
  title     = "Polycentric systems for coping with collective action and global
               environmental change",
  author    = "Ostrom, E.",
  journal   = "Glob. Environ. Change",
  publisher = "Elsevier BV",
  volume    =  20,
  number    =  4,
  pages     = "550--557",
  month     =  oct,
  year      =  2010,
  doi = "10.1016/j.gloenvcha.2010.07.004"
}

@ARTICLE{Smith2022-ds,
  title     = "Review and viability of a {Dyson} swarm as a form of {Dyson}
               sphere",
  author    = "Smith, J.",
  journal   = "Phys. Scr.",
  publisher = "IOP Publishing",
  volume    =  97,
  number    =  12,
  pages     = "122001",
  year      =  2022,
  doi = "10.1088/1402-4896/ac9e78"
}

@ARTICLE{Bracewell1960-bp,
  title     = "Communications from superior galactic communities",
  author    = "Bracewell, R. N.",
  journal   = "Nature",
  publisher = "Springer Science and Business Media LLC",
  volume    =  186,
  number    =  4726,
  pages     = "670--671",
  month     =  may,
  year      =  1960,
  doi = "10.1038/186670a0"
}

@ARTICLE{Xia2022-nw,
  title     = "Global food insecurity and famine from reduced crop, marine
               fishery and livestock production due to climate disruption from
               nuclear war soot injection",
  author    = "Xia, L. and Robock, A. and Scherrer, K. and others",
  journal   = "Nat. Food",
  publisher = "Springer Science and Business Media LLC",
  volume    =  3,
  number    =  8,
  pages     = "586--596",
  month     =  aug,
  year      =  2022,
  doi = "10.1038/s43016-022-00573-0"
}

@ARTICLE{Daly2023-dart,
  title     = "Successful kinetic impact into an asteroid for planetary defence",
  author    = "Daly, R. T. and Ernst, C. M. and Barnouin, O. S. and others",
  journal   = "Nature",
  publisher = "Springer Science and Business Media LLC",
  volume    =  616,
  number    =  7957,
  pages     = "443--447",
  month     =  apr,
  year      =  2023,
  doi = "10.1038/s41586-023-05810-5"
}

@MISC{Colossal2024-bv,
  title        = "Colossal BioVault",
  author       = "{Colossal Biosciences}",
  howpublished = "\url{https://colossalfoundation.org/project/species-biobank/}",
  note         = "Accessed: 2026-4-7"
}

@MISC{NOAA-GML-lz,
  title        = "Chlorofluorocarbons ({CFCs})",
  author       = "{NOAA Global Monitoring Laboratory}",
  howpublished = "\url{https://gml.noaa.gov/hats/publictn/elkins/cfcs.html}",
  year         =  2025,
  note         = "Accessed: 2025-6-16"
}

\end{document}